\documentstyle[aps,prl,multicol,epsf]{revtex}
\begin{document}

\begin{multicols}{2}

\noindent{\large\bf Comment on ``Nonuniversal Exponents in Interface Growth''}

\bigskip

In a recent Letter, Newman and Swift\cite{tim} made an interesting suggestion
that the strong-coupling exponents of the Kardar-Parisi-Zhang\cite{kpz} (KPZ)
equation reported previously\cite{forr90,halp} may not be universal, 
but rather depend on the precise form of the noise distribution. 
The purpose of this Comment is to show that their numerical findings 
can be attributed to a percolative effect instead of a portentous 
breakdown of universality. 

The interface growth model of Ref. 1 can be equivalently formulated as
\begin{mathletters}
\begin{eqnarray}
\tilde h_i(t)&=&h_i(t)+\xi_i(t),
\\
h_i(t+1)&=&\min_{j\in S_i}\{\tilde h_j(t)\},
\end{eqnarray}
\end{mathletters}
where the minimum in (1b) is taken over the set $S_i$ which
includes site $i$ and all its 
nearest neighbors on a $d$-dimensional hypercubic lattice.
In this form, the height variable $h_i(t)$ can be readily taken as the
ground state energy of a directed polymer on a $d+1$ dimensional
lattice with the upper end fixed at $(i,t)$.\cite{halp}

In Ref. 1, the noise (or random potential) term $\xi_i(t)$ in
Eq. (1a) was drawn from a distribution,
\begin{equation}
p_\alpha(\xi)={1+\alpha\over 2}(1-|\xi|)^\alpha,\qquad 
-1\leq\xi\le 1.
\end{equation}
It was observed numerically that, as the parameter $\alpha$
decreases towards its limiting value $-1$, the exponent $\beta$
which characterizes the scaling of the interface width
$W(t)\equiv (\langle h^2\rangle -\langle h\rangle^2)^{1/2}\sim t^\beta,$
appears to decrease dramatically, particularly in high dimensions ($d=3,4$).
This behavior led Newman and Swift to suggest that
the universality hypothesis is broken in the KPZ equation.

Here we give an alternative interpretation of the slow growth of $W(t)$
over the time interval they investigated. 
In the limit $\alpha=-1$, the distribution (2)
reduces to a discrete one where $\xi=\pm 1$ with equal probability.
For $d>1$, we found\cite{cct} that the $\xi=-1$ sites percolate in the sense 
defined by (1b). Consequently, $W(t)$ saturates to a 
constant after an initial transient, yielding $\beta=0$. 
(The directed percolation threshold $p_c$ equals 0.539 and 0.265 for
$d=1$ and 2, respectively, which explains why the ``nonuniversal''
behavior is absent in $d=1$.)
In the directed polymer language, this corresponds to a finite density of 
paths going through the minimum energy sites $\xi=-1$ and they all yield the
same ground state energy $h_i(t)=-t$. The ground-state degeneracy is lifted
when $\alpha>-1$. However, for $\alpha$ close
to $-1$, the effectively competing paths are those which utilize only
the bottom part of the distribution (2), i.e., those going through
sites with $\xi$ close to $-1$. Therefore, the effective noise
fluctuation is much weaker than what appears to be from (2).
This, plus an ``intrinsic width'' effect (i.e., $W$ remains finite
even for $\alpha=-1$), is the origin of the observed behavior.

\begin{figure}
\narrowtext
\epsfxsize=\linewidth
\epsffile{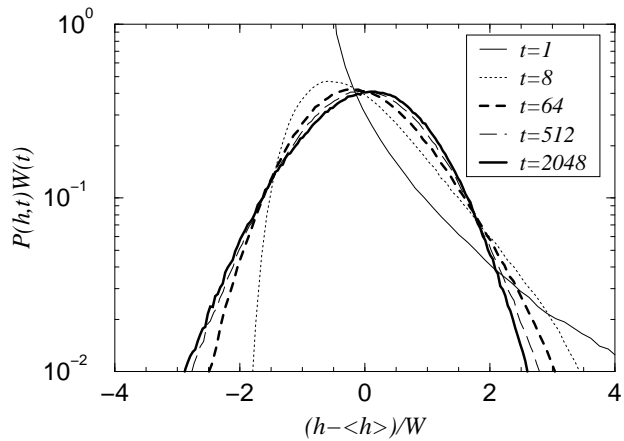}
\caption{Normalized distribution of surface height at different times
for $d=2$ with $2000^2$ sites. Here $\alpha=-1/2$.
}
\label{fig1}
\end{figure}

A quantitative analysis of the crossover to the asymptotic scaling
can be carried out by examining the evolution of the
surface-height distribution[5]. 
In the scaling regime, one expects the normalized distribution
to converge to a limiting form. From Fig. 1 we see that
such convergence is rather slow for $d=2$ and $\alpha=-0.5$.
The slow-convergence to the asymptotic distribution (which we believe
to be universal) is more pronounced for smaller values of $\alpha$
(weaker effective noise) or in higher dimensions
(diminishing directed percolation threshold $p_c$). Details of
our analysis, including calculations on the Berker lattice,
will be published elsewhere[5].

\bigskip
{\obeylines
\noindent Hugues Chat\'e
CEA --- Service de Physique de l'Etat Condens\'e
Centre d'Etudes de Saclay
91191 Gif-sur-Yvette, France
\bigskip
\noindent Qing-Hu Chen and Lei-Han Tang
Department of Physics
Hong Kong Baptist University
Kowloon Tong, Hong Kong
}


\end{multicols}

\end{document}